\documentclass[twocolumn,showpacs,prb,footinbib,a4paper,superscriptaddress,floatfix]{revtex4}

\usepackage{amssymb,amsmath}
\usepackage{graphics}
\usepackage{dcolumn}
\usepackage{bm}
\usepackage{epsfig}

\newcommand{\GF}{{\rm Green's function}\ }
\newcommand{\GFs}{{\rm Green's functions}\ }
\newcommand{\SE}{{\rm self-energy}\ }
\newcommand{\SEs}{{\rm self-energies}\ }
\newcommand{\NE}{{\rm non-equilibrium}\ }
\newcommand{\MB}{{\rm MB}}
\newcommand{\inter}{{\rm int}}
\newcommand{\summ}{\bullet\hspace{-3.6mm}\sum}
\newcommand{\summline}{\bullet\hspace{-2.9mm}\sum}
\newcommand{\tr}{{\rm Tr}}

\usepackage[usenames,dvipsnames]{color}
\definecolor{MyOrange}{rgb}{1.0,0.5,0}
\definecolor{MyPurple}{rgb}{0.5,0,1}

\begin{document}

\title{Non-equilibrium quantum transport in fully interacting single-molecule
nanojunctions}

\author{H. Ness}
\email{herve.ness@york.ac.uk}
\author{L. K. Dash}

\affiliation{Department of Physics, University of York, Heslington, York YO10 5DD,
UK}
\affiliation{European Theoretical Spectroscopy Facility (ETSF)}

\date{\today}

\begin{abstract}
  Using \NE Green's functions, we derive a formula for the electron current through
  a lead-molecule-lead nanojunction where the interactions are not restricted to the
  central region, but are spread throughout the system, including the leads and the
  lead-molecule interfaces.
  The current expression consists of two sets of terms. The first set corresponds to 
  a generalized Meir and Wingreen expression where
  the leads' self-energies are renormalized by the interactions
  crossing at the molecule-lead contacts. The second set corresponds to inelastic
  scattering events in the leads arising from any arbitrary interaction, including electron-electron
  and electron-phonon coupling, treated beyond mean-field approximations. 
  Using different levels of approximation, we are able to recover well-known
  expressions for the current.
  We also analyse how practical calculations can be performed with 
  our formalism  by using the new concept of
  generalized embedding potentials.
\end{abstract}
 
\pacs{71.38.-k, 73.40.Gk, 85.65.+h, 73.63.-b}

\maketitle

\section{Context}
\label{sec:intro}

Developing a theory for the \NE electronic quantum transport through nanoscale 
junctions is a challenging task, especially when thinking in terms of 
applications for nanoscale electronics.
Electronic transport through nanojunctions (single-molecule junctions, for 
example) exhibits many important
new features in comparison with conduction through macroscopic
systems. This leads to promising new applications in single-molecule
electronics. 
In particular, interactions such as Coulomb
interactions between the electrons and scattering from localized
atomic vibrations are critically important.  

Having a simple expression for the current (or the
conductance) of a nanoscale object connected to terminals is most
useful.  This is provided by the Landauer formula \cite{Landauer:1970}
which
describes the current in terms of local properties (transmission coefficients) 
of a finite central region $C$ and the distribution functions of the
electron reservoirs connected to this region $C$.  However, 
the original Landauer formulation deals only with non-interacting
electrons.  It has been used with success in conjunction with
density-functional theory calculations for realistic nanoscale
systems
\cite{Hirose:1994,
Taylor:2001,Brandbyge:2002,
Xue:2003,
Thygesen:2003}
since DFT maps the many-electron interacting system onto an 
effective single-particle problem.
However there are many cases when such a single-particle approach becomes
questionable \cite{Vignale:2009,Ness:2010}.
The Landauer formula has been built upon by Meir and Wingreen
\cite{Meir:1992} to extend the formalism to a central scattering
region containing interactions by using the non-equilibrium Green's functions
formalism.
Other generalizations of Landauer-like approaches to include interactions and inelastic
scattering in the region $C$ have been developed 
\cite{Bonca:1995,Ness:1999,
Zitko:2003,
Ferretti:2005a}.
However, in real systems the interaction is not confined to the central region but
exists throughout the system.
Accounting for the interaction along the whole system is vital \cite{Stefanucci:2004a, Vignale:2009,Myohanen:2010}. 

In this paper, we provide a complete description which generalizes the Meir
and Wingreen formalism to systems where
interactions exist throughout the system, as well as at the interfaces
between the central region and the electrodes.  Since the choice of the location of these
interfaces is purely arbitrary, and since the interactions exist everywhere, 
our approach is formally identical to a partition-free scheme \cite{Cini:1980,Stefanucci:2004a}.
While keeping the approach 
of the original work of Meir and Wingreen \cite{Meir:1992}, we derive the
most general expression of the current for the fully interacting
system.  From this, we recover all previously derived transport
expressions or corrections when introducing the appropriate level of
approximation for the interaction.
Our formalism also leads to the generalization of the concept of embedding potentials 
when the interaction crosses at the boundaries. It therefore provides an alternative
way of introducing open boundary conditions with interaction in finite-size systems.

The paper is organized as follows.
In Section \ref{sec:NEtransport}, we 
provide the generalised current formula for fully interacting systems.
We describe our model in Section \ref{sec:model}
and derive the current expression in Section \ref{sec:IL}, 
with full details of the calculations provided in Appendix \ref{app:derivationIL}.
The connections between more conventional results: the current at equilibrium, the current
formula of Meir and Wingreen and others
are given in Sections \ref{sec:ILatequi} to \ref{sec:ILwithdistribfnc}.
In section \ref{sec:sigmaLC_Hartree} we describe how to apply our formalism in a specific case of
interaction crossing at the contacts.
Finally, we conclude and discuss extension of our work in Section \ref{sec:conclu}.

\section{Non-equilibrium quantum transport}
\label{sec:NEtransport}

\subsection{The model}
\label{sec:model}

The system consists of two electrodes, left $L$ and right $R$, which connect a
central region $C$ via coupling matrix elements. The
interaction, which we specifically leave undefined
(e.g. electron-electron or electron-phonon), is spread over the entire system
and crosses at the interfaces between the $L(R)$ and $C$ regions.
We use different labels for the quantum states on each side of these
interfaces:
$\{\lambda,\lambda'\},\{n,m\},\{\rho,\rho'\}$ are used to represent 
the complete and orthogonal set of states for 
the $L,C$ and $R$ regions respectively.
We also use a compact notation for the matrix elements $M$ 
of Green's functions ($g,G$), the self-energies ($\Sigma$) and coupling to the leads ($V$), where
$M_C$ represents the matrix elements $M_{nm}$ in the region $C$, 
$M_{LC}$ for $M_{\lambda m}$, $M_{CL}$ for $M_{n \lambda'}$, 
$M_{RC}$ for $M_{\rho m}$, $M_{CR}$ for $M_{n \rho'}$,
$M_L$ for $M_{\lambda \lambda'}$, and $M_R$ for $M_{\rho \rho'}$.

The complete derivation of the current expressions for the fully interacting lead-central region-lead
junctions relies on only two assumptions:
the many-body effects of the interacting particles are well described by
self-energies $\Sigma^{\rm MB}$ in the one-particle Green's functions $G$,
and there is no direct coupling or interaction between the states of the $L$ and $R$ regions:
the only interaction between the leads is mediated by the region $C$, there is no
direct coupling 
i.e. $\Sigma^{\rm MB}_{\lambda \rho (\rho' \lambda')}=0$ \cite{footnote_SigmaLR}.

\begin{figure}
  \center
  \includegraphics[width=\columnwidth]{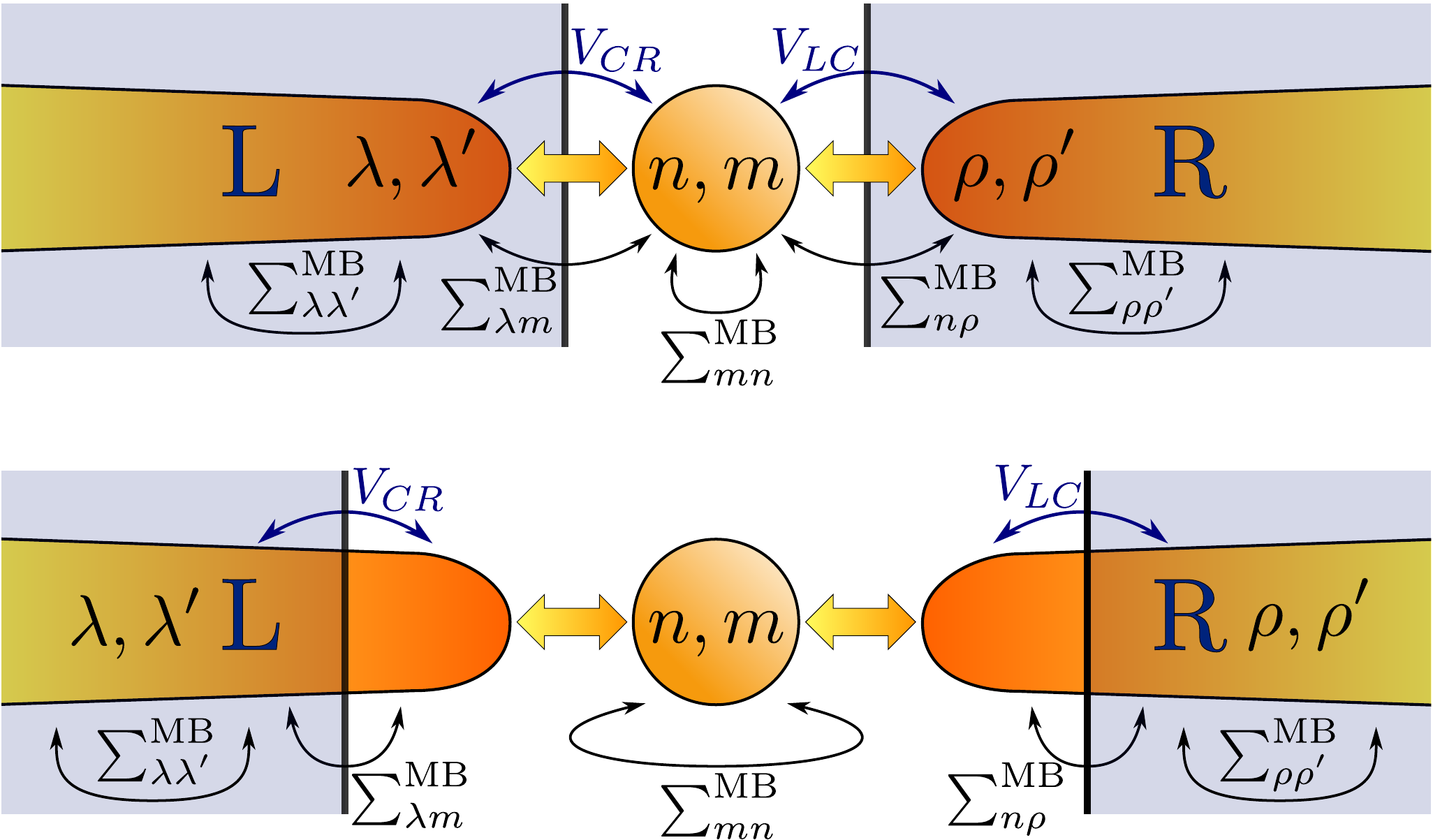}
  \caption{Schematic representation of a central scattering 
region $C$ connected to the left $L$ and right $R$ electrodes, with respective quantum-state labels $\{\lambda\},\{n\},\{\rho\}$ for the three $L,C,R$ subspaces.
Interactions are given by
the coupling of the region $C$ to the $L(R)$ electrode $V_{LC/CL}$ ($V_{RC/CR}$),
and by the many-body effects $\Sigma^{\rm MB}$ within all regions as well as
across the $LC$ and $CR$ interfaces. 
Top: interfaces are arbitrarily placed at the contact between the 
scatterer (a molecule) and the leads. They 
cannot be considered as being at equilibrium, being each in a region of 
strong spatial variation of the current and potential drop.
Bottom:  interfaces are now well inside the $L,R$ regions ($C$ region is now
the so-called extended molecule), and are at local quasi-equilibrium.}
\label{fig:system}
\end{figure}

\subsection{The non-equilibrium current}
\label{sec:IL}

The location of the interfaces $LC$ and $CR$ is
arbitrary (Fig.~\ref{fig:system}), and chosen purely for mathematical convenience, 
as the interaction spreads throughout the system. 
We include such interfaces to make the connection between our results and
other previously derived expressions within
the partitioning scheme.

From the continuity equation $\nabla \vec{j} + \partial_t n =0$, we 
write the current through the interface between the $L$
and $C$ regions. 
The current flowing at 
the $LC$ interface is given by
$I_L(t)	= - e \partial_t \langle \hat{N}_L(t) \rangle$, 
where $\langle \hat{N}_L(t) \rangle$ is the number of electrons
in the $L$ region, and is related to the lesser Green's function as
$ \langle \hat{N}_L(t) \rangle = \sum_\lambda -{\rm i} G^<_{\lambda\lambda}(t,t)$. 
From the equations of motion \cite{Keldysh:1965,Danielewicz:1984} 
obeyed by the Green's functions on the Keldysh time-loop contour $C_K$, 
we obtain the current $I_L$ as 
\begin{equation}
\label{eq:ILafterEOM}
I_L(t) = \frac{e}{\hbar} \tr_\lambda 
\left[ (\Sigma G)^<(t,t) - (G \Sigma)^<(t,t) \right]
\end{equation}
From the rules of analytical continuation on $C_K$
(see Appendix \ref{app:analyticalcontinuation}), we find that 
\begin{equation}
\begin{split}
(\Sigma G)^< = \Sigma^{\MB,<} G^a +  ( V_{LC} + \Sigma^{\MB,r}) G^< , \\
(G \Sigma)^< =  G^< (V_{CL} + \Sigma^{\MB,a} ) + G^r\Sigma^{\MB,<} .
\end{split}
\end{equation}
There are no lesser (greater) components for $V_{LC}$ since its time
dependence is local $V_{LC}(t,t') = V_{LC}(t) \delta(t-t')$.

In the steady state, all double-time quantities $X(t,t')$ depend only on the time difference
$X(t-t')$. The steady state current is given after Fourier transform by
using $(\Sigma G)^<(t,t) \rightarrow \int {\rm d}\omega / 2\pi \Sigma(\omega) G(\omega)$.
To obtain the current, we need to calculate the following trace:
\begin{equation}
\label{eq:Trace_in_ILafterEOM_omega}
\begin{split}
& \tr_\lambda \left[\dots \right] =
\sum_{\lambda,n,\gamma}   
	  V_{\lambda n} G^<_{n \lambda}(\omega) - G^<_{\lambda n}(\omega) V_{n \lambda}  \\
&  	+ \Sigma^{\MB,<}_{\lambda \gamma}(\omega) G^a_{\gamma \lambda}(\omega) 
	+ \Sigma^{\MB,r}_{\lambda \gamma}(\omega) G^<_{\gamma \lambda}(\omega)  \\ 
&  	-  G^<_{\lambda \gamma}(\omega) \Sigma^{\MB,a}_{\gamma \lambda}(\omega) 
	-  G^r_{\lambda \gamma}(\omega) \Sigma^{\MB,<}_{\gamma \lambda}(\omega) ,
\end{split}
\end{equation}
where $\gamma$ runs only on the $L$ and $C$ regions, since 
$\Sigma^{\rm MB}_{\lambda \rho}=0$ (there is no direct coupling
between the $L$ and $R$ regions).
We then need to evaluate the \GFs matrix elements
$G^<_{n\lambda}, G^<_{\lambda n}, G^a_{n\lambda}$ and $G^r_{\lambda n}$,
and $G^{<,r,a}_{\lambda \lambda'}$ by using the Dyson equation 
$G^x_{ij}=g^x_{ij}+[g\Sigma G]^x_{ij}$ (with $x=r,a,<$ and $\{i,j\}$ the indices
for the corresponding matrix elements) 
and the rules of analytical continuation for the products (see Appendices
\ref{app:analyticalcontinuation} and \ref{app:derivationIL} for detail).

We find the following general expression for the current $I_L$ flowing 
through the left interface:
\begin{equation}
     \label{eq:ILfinal}
\begin{split}
 I_L & = \frac{e}{\hbar} \int \frac{d\omega}{2\pi} \\
& \tr_n\left[ G^r_C \tilde{\Upsilon}^l_{LC} + G^a_C
  (\tilde{\Upsilon}^l_{LC})^\dagger  + G^<_C(\tilde\Upsilon_{LC} -
  \tilde{\Upsilon}_{LC}^\dagger) \right] \\
+ & \tr_\lambda\left[\Sigma^{\MB,>}_{\lambda\lambda^\prime}
   G^<_{\lambda^\prime\lambda} -
   \Sigma^{\MB,<}_{\lambda\lambda^\prime} G^>_{\lambda^\prime\lambda} \right]
\end{split}
\end{equation}
where
\begin{equation}
\label{eq:Upsilons}
\begin{split}
\tilde\Upsilon_{LC}  & = \Sigma^a_{CL}\ \tilde{g}^a_L\ \Sigma^r_{LC} , \\
\tilde\Upsilon^\dag_{LC} & = \Sigma^a_{CL}\ \tilde{g}^r_L\ \Sigma^r_{LC} , \\
\tilde\Upsilon^l_{LC}  & = \Sigma^<_{CL} \left( \tilde{g}^a_L - \tilde{g}^r_L \right) \Sigma^r_{LC}
+ \Sigma^r_{CL}\ \tilde{g}^<_L\ \Sigma^r_{LC} .
\end{split}
\end{equation}
By definition $\Sigma_{LC}(\omega)= V_{LC} + \Sigma^\MB_{LC}(\omega)$, and similarly for the $CL$
components.
$\tilde{g}^{r,a}_L(\omega)$ are the \GFs of the region $L$ renormalised by the interaction
{\em inside} that region:
$(\tilde{g}^{r/a}_L)^{-1} = ({g}^{r/a}_L)^{-1} - \Sigma^{{\rm MB},r/a}_L$
where all quantities are defined only in the subspace $L$.

There are two contributions to $I_L$: the first trace is a generalisation 
of the Meir and Wingreen expression \cite{Meir:1992} to the cases where the interactions exist
within the three $L,C,R$ regions as well as in between the regions.
The different quantities $\Upsilon_{LC}$ are related to the generalised embedding
potentials (i.e.~lead self-energies) with interaction crossing at the $LC$ and $CR$
interfaces (see end of Appendix \ref{app:derivationIL}). 
The second trace in Eq.~(\ref{eq:ILfinal}) is related to inelastic effects 
involving a sum over the states of the $L$ region. 
Although the $L$ region is semi-infinite by definition, an appropriate choice of the
location of the $LC$ interface reduces the summation.
For a closed system at equilibrium, the trace 
$\tr\left[\Sigma^> G^< - \Sigma^< G^>\right]$
is zero simply because the system obeys the detailed balance equation:
$\Sigma^> G^< = \Sigma^< G^>$.
For all other conditions, if the $LC$ interface is located deep enough in the $L$ electrode,
the system is locally at quasi-equilibrium, and hence the trace vanishes (see below).

An expression similar to Eq.~(\ref{eq:ILfinal}) can be obtained for the current 
$I_R$ flowing at the right {\em CR} interface by swapping the
index $L$~$\leftrightarrow$~$R$ and using the current conservation condition
$I_L+I_R=0$.

\subsubsection*{Different Green's functions}

Finally we need to know, for practical calculations, 
the different \GFs in all three regions. 
To evaluate the currents $I_{L,R}$, we need the \GFs $G^{a/r,<}_C$ and $G^{a/r,<}_{L,R}$. 


We find for $G^r_C = \langle n\vert G^r \vert m\rangle$
\begin{equation}
\label{eq:GrC}
\begin{split}
G^r_C(\omega) & = {g}^r_C + {g}^r_C\ \Sigma^{\MB,r}_C\ G^r_C + {g}^r_C\ \tilde Y^r_{L+R}\ G^r_C \\
      & = \left[ ({g}^r_C(\omega))^{-1} - \Sigma^{\MB,r}_C(\omega) -  \tilde Y^r_{L+R}(\omega)
      \right]^{-1} 
\end{split}
\end{equation}
where $\tilde Y^r_{L+R}$ is the sum of the generalised leads'
self-energies $\tilde Y^r_\alpha$ ($\alpha=L,R$) defined as
$\tilde Y^r_\alpha= (\Sigma_{C\alpha} \tilde g_{\alpha} \Sigma_{\alpha C})^r$.

We find for $G^r_R = \langle \rho \vert G^r \vert \rho' \rangle$ that
\begin{equation}
\label{eq:GrR}
G^r_{\rho \rho'} = \tilde{g}^r_{\rho \rho'} 
+ \tilde{g}^r_{\rho \rho_1}\ \tilde{\tilde Y}^{r}_{C,\rho_1\rho_2} G^r_{\rho_2 \rho'} 
\end{equation}
where $\tilde{\tilde Y}^{r}_{C}$ is the embedding potential arising from
the central region $C$. It is defined in the right region $R$ as follows:
\begin{equation}
\label{eq:Cregion_embed_pot1}
\tilde{\tilde Y}^{r}_{C,\rho_1\rho_2}(\omega) 
= \Sigma^r_{\rho_1 m}(\omega)\ \tilde{\tilde{g}}^r_{ml}(\omega)\ \Sigma^r_{l \rho_2}(\omega) ,
\end{equation}
with $\Sigma^r_{RC}=V_{RC}+\Sigma^{\MB,r}_{RC}$ (similarly for
$\Sigma^r_{CR}$). $\tilde{\tilde{g}}^r_C$ is a retarded \GF of the
region $C$ renormalized by the interaction inside the central region $C$ and
by the embedding potential of the left region $L$ only:
\begin{equation}
\label{eq:Cregion_embed_pot2}
\tilde{\tilde{g}}^r_C = \left[ (\tilde{g}^r_C(\omega))^{-1}  - \tilde Y^r_L(\omega) \right]^{-1} .
\end{equation}

The form of these equations hold for the \GF $G^r_L$ as well as for the advanced \GFs $G^a_{L,R}$.

We finally get for $G^<_C = \langle n\vert G^< \vert m\rangle$:
\begin{equation}
\label{eq:GlessC}
G^<_C = G^r_C\ \left( \Sigma^{\MB,<}_C + \tilde Y^<_{L+R} \right) G^a_C,
\end{equation}
with 
$Y^<_{L+R}(\omega) = \sum_{\alpha=L,R} 
\left( \Sigma_{C\alpha} \tilde g_{\alpha} \Sigma_{\alpha C} \right)^<$.
The rules of analytical continuation need to be applied to the products 
$\left( \Sigma_{C\alpha}(\omega) \tilde g_{\alpha}(\omega) \Sigma_{\alpha C}(\omega) \right)^{<,r,a}$
to get the full expansion of the generalised embedding potentials.

\subsection{The current at equilibrium}
\label{sec:ILatequi}

One of the obvious checks to perform is that there is no net current at
the $LC$ and $CR$ interfaces at equilibrium.  Considering the equation
for $I_L$ given by Eq.(\ref{eq:ILfinal}), we have already shown that
the trace ${\rm Tr}_{\lambda} [...]$  vanishes at equilibrium
because of the detailed balance principle.  Now we have to
prove the same for the trace ${\rm Tr}_{n} [...]$ in
Eq.(\ref{eq:ILfinal}).  For this we use the procedure which consists
of introducing \NE distribution functions (see Section \ref{sec:ILwithdistribfnc} below).
Since at equilibrium all distributions are equal to the Fermi
distribution $f^{\rm eq}$, we end up, after long but trivial manipulation
of Eq.(\ref{eq:ILfinal}), with
\begin{equation}
\label{eq:def_spectralfncs1}
\begin{split}
{\rm Tr}_{n} [...]^{\rm eq} = 
{\rm Tr}_{n} \left[ 
G^r_C \Sigma^r_{CL} \left( f^{\rm eq} \tilde{g}^r_L - \tilde{g}^r_L f^{\rm eq} \right) \Sigma^r_{LC} \right. \\
\left. - G^a_C \Sigma^a_{CL} \left( f^{\rm eq} \tilde{g}^a_L - \tilde{g}^a_L f^{\rm eq} \right) \Sigma^a_{LC}
\right] ,
\end{split}
\end{equation}
which after further manipulation (using complex-conjugate relations between
\GFs and \SEs) can be shown to be equal to zero. Hence, as expected, the current
$I_L$ from Eq.(\ref{eq:ILfinal}) vanishes at equilibrium.

\subsection{Recovering the Meir and Wingreen current formula}
\label{sec:MeirWin}

For systems where there are interactions only within $C$, we have 
$\Sigma^{\MB}_{nm} \ne 0$ and $\Sigma^{a/r}_{LC/CL}=V_{LC/CL}$.
Then $\tilde{g}^x_{\alpha} \equiv {g}^x_{\alpha}$, and
$\tilde\Upsilon_{LC} = V_{CL}\ {g}^a_L\ V_{LC}$,
$\tilde\Upsilon^l_{LC} = V_{CL}\ {g}^<_L\ V_{LC} = -
(\tilde\Upsilon^l_{LC})^\dag$, and
Eq.~(\ref{eq:ILfinal}) can be recast as
\begin{equation}
\label{eq:IL_MeirWingreen}
I_L = \frac{{\rm i}  e}{\hbar} \int \frac{{\rm d}\omega}{2\pi}\
{\rm Tr}_{n} \left[ f_L (G^r_C - G^a_C) \Gamma_L + G^<_C \Gamma_L \right] ,
\end{equation}
with ${\rm i} f_L \Gamma_L=V_{CL}\ {g}^<_L\ V_{LC}$
and
${\rm i}\Gamma_L=V_{CL} ({g}^a_L-g^r_L) V_{LC}$.
Hence we recover the result of Meir and Wingreen \cite{Meir:1992}.

Going one step further, we consider interaction within the $L$ and $R$ regions
as well. The current $I_L$ in Eq.(\ref{eq:ILfinal}) takes
then the form of the Meir and Wingreen expression
Eq.(\ref{eq:IL_MeirWingreen}), with renormalised escape rates
$\tilde\Gamma_L$, i.e.\  ${\rm i}\tilde\Gamma_L=V_{CL}\
(\tilde{g}^a_L-\tilde{g}^r_L)\ V_{LC}$ and
$\tilde\Upsilon^l_{LC} \equiv {\rm i} \tilde{f}_L \tilde\Gamma_L(\omega) 
= V_{CL}\ \tilde{g}^<_L\ V_{LC} $.
The interactions within the leads renormalise the coupling at
the contacts $\tilde\Gamma_{L}$.  Note that we have
allowed for a renormalised distribution function $\tilde{f}_L$ in the definition of
$\tilde\Upsilon^l_{LC}$.  The distribution of the left lead
$\tilde{f}_L$ has the same form as the Fermi distribution function,
but depending on the approximation chosen for the interaction
$\Sigma^{\MB}_L$, the corresponding Fermi level may also need renormalization.

\subsection{Transport with interaction on the (TD)DFT level}
\label{sec:ILatDFTlevel}

We consider cases where the interaction is spread throughout the entire
system, and are treated at the
level of density-functional theory (DFT).  The exchange and
correlation effects for interacting electrons are
given by an effective potential $v_{xc}(r,t)$ obtained from an $xc$
action functional of the electron density. To this potential corresponds
an effective self-energy, local in both space and time \cite{Stefanucci:2004a,Vignale:2009}.
This forms a class of self-energies, where 
$\Sigma^{\MB}(\tau,\tau')=\hat\Sigma^{\MB}(\tau)\delta(\tau-\tau')$ 
cannot have lesser or greater 
components, since the times $\tau$ and $\tau'$ must be on the 
same time-loop branch.
With no lesser and greater components for $\Sigma^{\MB}$,
the trace ${\rm Tr}_{\lambda}[...]$ in Eq.(\ref{eq:ILfinal}) simply vanishes. 
We are thus left with
\begin{multline}
\label{eq:ILBurke}
I_L = \frac{e}{\hbar} \int \frac{{\rm d}\omega}{2\pi}\ {\rm Tr}_{n}
\left[ G^r_C\ \tilde\Upsilon^l_{LC}
  + G^a_C\ (\tilde\Upsilon^l_{LC})^\dag \right. \\
  \left. + G^<_C \left( \tilde\Upsilon_{LC} - \tilde\Upsilon^\dag_{LC}\right)
\right] ,
\end{multline}
where
$\tilde\Upsilon^l_{LC}  = \Sigma_{CL}\ \tilde{g}^<_L\ \Sigma_{LC}$ ,
$(\tilde\Upsilon^l_{LC})^\dag = - \Sigma_{CL}\ \tilde{g}^<_L\ \Sigma_{LC}$,
$\tilde\Upsilon^\dag_{LC}=\Sigma_{CL}\ \tilde{g}^r_L\ \Sigma_{LC}$,
and $\Sigma = V +v_{xc}$ ($V$ has only $V_{\alpha C/C \alpha}$ components,
and $v_{xc}$ has local static or dynamic
components $v_{xc, \{\lambda, n, \rho\}}$ for DFT or
time-dependent DFT calculations respectively) \cite{footnote_NLvxc}
Hence we recover a Meir-and-Wingreen-like expression for the current
with renormalised $\tilde\Gamma_L$.
The potential $v_{xc}$ is spread throughout
the system and inside the leads \cite{Stefanucci:2004a,Vignale:2009}.
Hence Eq.~(\ref{eq:ILBurke}) formally confirms the necessity of including 
the potential drop due to $v_{xc}$ in the linear-response regime \cite{Koentopp:2008}.

One should note that our formalism includes all other cases with other kind of interactions 
(electron-phonon) confined only in the central region 
\cite{Frederiksen:2004, Galperin:2004b,
Sergueev:2005, 
  Yamamoto:2005, 
Dash:2010,Dash:2011}).  It also includes other
kind of electron-hole excitations which can be present in the 
leads \cite{Galperin:2006b}; and provides a way to treat systems with electron-hole excitations
crossing at the contacts between the central region and the leads.

\subsection{The current in terms of distribution functions and spectral densities}
\label{sec:ILwithdistribfnc}

We now discuss when the second trace in Eq.~(\ref{eq:ILfinal}) vanishes. 
We introduce the \NE distributions $f^<(\omega)$ obtained from the generalised
Kadanoff-Baym ansatz \cite{Lipavski:1986}
$X^<(\omega) = f^<(\omega) X^a(\omega) - X^r(\omega) f^<(\omega)$
for a \GF or a \SE $X$. 
We define
\begin{equation}
\label{eq:NEdistrib}
\begin{split}
\tilde{g}^<_L & = f^{0<}_L \tilde{g}^a_L - \tilde{g}^r_L f^{0<}_L , \\
G^<_C & = f^<_C G^a_C - G^r_C f^<_C , \\
\Sigma^{<}_{LC} & = f^{\inter <}_{L} \Sigma^{a}_{LC} - \Sigma^{r}_{LC} f^{\inter <}_{C} , \\
\Sigma^{<}_{CL} & = f^{\inter <}_{C} \Sigma^{a}_{CL} - \Sigma^{r}_{CL} f^{\inter <}_{L}
\end{split}
\end{equation}
(with $f^{0<}_L$ the Fermi distribution $f_L$ of the $L$ region). We 
rewrite Eq.~(\ref{eq:ILfinal}) as follows
\begin{equation}
\label{eq:app_IL_and_NEdistrib}
\begin{split}
I_L = \frac{e}{\hbar} \int & \frac{{\rm d}\omega}{2\pi}\
  {\rm Tr}_n \left[ 
\delta G^<_C\ \Sigma^a_{CL} \left( \tilde{g}^a_L - \tilde{g}^r_L
\right)  \Sigma^r_{LC} \right] \\
 + & {\rm Tr}_\lambda \left[ \left(
\Sigma^r_{LC} G^r_C \Sigma^r_{CL} - \Sigma^a_{LC} G^a_C \Sigma^a_{CL} \right) 
\delta\tilde{g}^<_L \right]  \\
 + & (2\pi)^2  {\rm Tr}_\lambda \left[\delta f^<_L\  A^\Sigma_L(\omega) A^G_L(\omega) \right],
\end{split}
\end{equation}
with
$\delta{g}^<_L  = \delta f^{0<}_L\ \tilde{g}^a_L - \tilde{g}^r_L\ \delta f^{0<}_L$, 
$\delta G^<_C  = \delta f^<_C\ G^a_C - G^r_C\ \delta f^<_C$.
The differences of distributions are 
$\delta f^{0<}_L  = f^{0<}_L - f^{\inter <}_L$,
$\delta f^{<}_{L,C}   = f^<_{L,C} - f^{\inter <}_{L,C}$, and the spectral functions are 
$A^X_\alpha=(X^a_\alpha-X^r_\alpha)/2\pi{\rm i}$.

At equilibrium all distributions are equal to the Fermi distribution, all
$\delta f=0$ and $I_L=0$ as expected.
For interactions localised in $C$ only, we again recover
Eq.(\ref{eq:IL_MeirWingreen}) by noticing that
$\Sigma^{a/r}_{LC(CL)}=V_{LC(CL)}$, $A^\Sigma_L=0$ and
$\tilde{g}_L=g_L$.
Furthermore, when the $LC$ interface is located well inside the $L$ region, the states $\lambda$ on the left side 
of the interface are at their local equilibrium. Hence the corresponding distributions
are equal to the local Fermi distribution and  
$\delta f^{0<}_L  = \delta f^{<}_L = 0$; and thereforce the traces $\tr_\lambda[...]$ in 
Eq.~(\ref{eq:app_IL_and_NEdistrib}) and in Eq.~(\ref{eq:ILfinal}) vanish (QED). 

 The current expression reduces then to
\begin{equation}
I_L = \frac{e}{\hbar} \int \frac{{\rm d}\omega}{2\pi}\
 {\rm Tr}_n \left[ 
\delta G^<_C\ \Sigma^a_{CL} \left( \tilde{g}^a_L - \tilde{g}^r_L \right)  \Sigma^r_{LC} \right]
\end{equation}
which is just another way to express the ${\rm Tr}_n [...] $ in Eq.(\ref{eq:ILfinal}).

\subsection{An example of crossing interaction}
\label{sec:sigmaLC_Hartree}

We now give a brief description of how to implement our formalism for a specific case. We consider a
single-molecule junction in the presence of electron-vibron interaction inside the central region and
crossing at one of the contacts. We use the following Hamiltonian for the central region
\begin{equation}
\label{eq:HCentral}
  H_C = \varepsilon_0 d^\dagger d + \omega_0 a^\dagger a +
  \gamma_0 (a^\dagger + a) d^\dagger d
\end{equation}
where one electronic level $\varepsilon_0$ and one vibration mode 
of energy $\omega_0$ are coupled together via the coupling constant $\gamma_0$.  
The central region is coupled to the non-interacting $L$ and $R$ regions via hopping integrals $t_{0\alpha}$:
\begin{equation}
\label{eq:Vcoupling}
  V_{LC} + V_{CR} = \sum_{\alpha = L, R} t_{0\alpha} (c^\dagger_\alpha
  d + d^\dagger c_\alpha) .
\end{equation}
We also consider that the hopping of an electron from the $C$ to the $L$ region (and {\em vice versa})
can excite another vibration mode of energy $\omega_A$ via the coupling constant $\gamma_A$:
\begin{equation}
\label{eq:Vcrossing}
H_{LC} = \gamma_A (b^\dagger + b)(c^\dagger_L d + d^\dagger c_L) + \omega_A b^\dagger b . 
\end{equation}
This
model can be understood as a lowest-order expansion of the hopping integral 
$t_{0L}(X)= t_{0L}+t'_{0L} X$ between the $C$ and $L$ regions in terms of the relative position
$X=\sqrt{{\hbar}/{(2m_A\omega_A)}}(b^\dagger + b)$ of the region $C$ with respect to the region $L$.
The Hamiltonian $H_{LC}$ represents in this model the interaction crossing at the $LC$ interface.
The corresponding \NE \GFs and \SEs can be calculated at different orders of the interaction 
using conventional \NE techniques \cite{Dash:2010,Dash:2011}.

Since there is no other interaction inside the $L$ and $R$ regions, the
current expression is given by the first line of Eq.(\ref{eq:ILfinal}). We consider a mean-field
approach to treat the crossing interaction. This model
leads to the Hartree-like expressions for the self-energies at the $LC$ interface:  
\begin{equation}
\label{eq:Vcrossing_Hartree}
\Sigma^{\MB,r/a}_{LC}=-2\frac{\gamma_A^2}{\omega_A}{\rm i}\int\frac{{\rm d}\omega}{2\pi} G^<_{LC}(\omega)
\end{equation}
(similarly for $\Sigma^{\MB,r/a}_{CL} \propto \int {\rm d}\omega G^<_{CL}(\omega)$).

The closed expression for $G^<_{LC}$ \GF matrix elements are calculated from the corresponding Dyson
equations $G^<_{LC}=[g\Sigma G]^<_{LC}$.
There are no lesser and greater components for the self-energy $\Sigma^{\MB}_{LC}$ at the mean-field 
level, as we have explained in Section \ref{sec:ILatDFTlevel}. 
Hence Eq.(\ref{eq:Upsilons}) reduces to
\begin{equation}
\label{eq:Hartreenorm}
\begin{split}
\tilde\Upsilon_{LC}           & = \Sigma^a_{CL}\ {g}^a_L\ \Sigma^r_{LC} , \\ 
\tilde\Upsilon^\dag_{LC}      & = \Sigma^a_{CL}\ {g}^r_L\ \Sigma^r_{LC}, \\
\tilde\Upsilon^l_{LC}         & = \Sigma^r_{CL}\ {g}^<_L\ \Sigma^r_{LC}, \\
(\tilde\Upsilon^l_{LC})^\dag  & = - \Sigma^a_{CL}\ {g}^<_L\ \Sigma^a_{LC}, \\
\end{split}
\end{equation}
with $\Sigma^{r/a}_{LC}=t_{0L}+\Sigma^{\MB,r/a}_{LC}$.

One can see that the interaction crossing at the $LC$ interface induces a static 
(however bias-dependent) renormalisation of the nominal coupling $t_{0L}$ between the $L$ and $C$ 
regions. This \NE renormalisation induces bias-dependent modifications of the broadening of the 
spectral features of the $C$ region. It can also lead to new physical \NE effects in the 
current \cite{Ness:2012unpub}.

The effects of the crossing interaction can also be treated beyond the mean-field level by considering
a Fock-like dynamical self-energy \cite{Dash:2010,Dash:2011}. In general the effects of other interaction
(electron-electron) crossing at the $LC$ and/or $CL$ interfaces can be treated in a similar manner.

\section{Discussion and conclusion}
\label{sec:conclu}

We have derived a exact expression for
  the current through systems with interaction both within the $L,C,R$ regions and
  at the $LC$ and $CR$ interfaces. Our result,
  Eq.(\ref{eq:ILfinal}), is general, assuming that there are 
  no direct interactions between the leads; a condition that is physically sound, especially
  for single-molecule junctions where the spatial gap between the two
  electrodes is large enough.
The location of the $LC$ and $CR$ interfaces with respect to the physical realistic
  scatterer is arbitrary but, in practice, should be
  chosen conveniently for numerical calculations.
  When local
  quasi-equilibria are reached at the interfaces, a simpler expression for the current
  is obtained, since the local
  non-equilibrium distribution functions are equal to the corresponding Fermi distributions. 
The deviations $\delta f^{0<}$ and $\delta f^{<}$ represent a quantitative
tool to determine how far inside the leads the $LC/RC$ interfaces need to be
to reach local equilibrium.
Our formalism provides a formal justification of the concept of
  the extended molecule that is commonly used with the
  conventional partitioned scheme.
It also provides the correction terms needed to deal
  with interaction crossing at the contacts and when the contacts are not 
in their respective local (quasi) equilibrium.

In practice, the calculations should be performed self-consistently since
the various self-energies $\Sigma^{\rm MB}$ in the three regions and at the
interfaces are functionals of the all \GFs in all the system. This offers extra
degrees of freedom to perform non-fully self-consistent calculations, and test 
different levels of approximations for the interaction.
We have given an example of how such calculations can be performed for a specific
case in Section \ref{sec:sigmaLC_Hartree}.
We have also found that the current conservation conditions lead to an
important result for a fully-interacting system: a condition that
the many-body \SEs $\Sigma^\MB$ should satisfying in order to keep the conservation
$I_L+I_R=0$ (see Appendix \ref{app:currentconserv}).

In a broader context, our formalism introduces in a formal
  manner the concept of generalised embedding potentials to
  interacting cases.
Embedding methods provide the correct boundary conditions for solving the Schr\"odinger equation in a limited region of space, region I, automatically matching the solution on to the wavefunction in the rest of the system, region II, via the use of the
embedding potential \cite{Inglesfield:1981}. In the quantum transport community, the embedding potentials usually arise from the left and right leads to which the scatterer of interest is connected, and are commonly referred to as the lead self-energies.
In conventional non-interacting approaches, they are given by
  $\Sigma_\alpha(\omega) = V_{C \alpha} g_\alpha(\omega) V_{\alpha
    C}$.
In our formalism, when the interactions cross the $LC/CR$
  interfaces, we obtain a generalisation of the
  embedding potentials, defined as $\tilde Y_\alpha^x(\omega) =
  (\Sigma_{C\alpha}(\omega) \tilde g_{\alpha}(\omega) \Sigma_{\alpha
    C}(\omega))^x$.
  These generalised embedding potentials contain a double
  non-locality, in the sense that the many-body part of
  $\Sigma_{\alpha C}$ has a spatial extent different from that of the coupling
  matrix elements $V_{\alpha C}$.  Hence $\tilde Y_\alpha$ defines
  a buffer zone, contained between two surfaces whose separation is related 
  to the
  characteristic spatial range of the interaction \SE $\Sigma^\MB_{\alpha C}
  \equiv \Sigma^{\rm MB}(\vert \mathbf{x}_\alpha -\mathbf{x}_n \vert)$.
The generalised embedding potential provides a new alternative for 
introducing open boundary conditions with interaction within many-body finite size systems.

In summary therefore, we have introduced a new formalism for an accurate expression
for the electron current in fully interacting systems. The expression is general and
takes into account the fact that the interaction is crossing through the interface on
which the current-density is integrated. Numerical implementations of our formalism 
will enable us to study cases in which the long-range Coulomb interaction is not 
sufficiently screened between the central region and the electrodes to be neglected or approximated, 
or cases in which vibration excitations at the contacts play an important role in the 
transport properties.

\appendix

\section{Relationship and symmetry on the Keldysh contour}
\label{app:GFSEdefinitions}

The relations between the different components of the Green's
functions and self-energies on the Keldysh time-loop contour $C_K$ are given
by:
\begin{equation}
\begin{split}
X^r = X^{++}-X^{+-} &= X^{-+}-X^{--} \\
X^a = X^{++}-X^{-+} &= X^{+-}-X^{--} \\
X^{++}+X^{--} &= X^{+-}+X^{-+}     \\ 
X^{-+}-X^{+-} &= X^r-X^a ,
\end{split}
\label{eq:app_gendef}
\end{equation}
with $X^{\eta_1 \eta_2}(12)\equiv G^{\eta_1 \eta_2}(12)$ or
$\Sigma^{\eta_1 \eta_2}(12)$, and where $(i=1,2)$ is the composite index for space-time
location $(\mathbf{x}_i,t_i)$ and $\eta_i$ is the index of the Keldysh time-loop contour 
$C_K$ branch ($+$ forward time arrow, $-$ backward time arrow) on which the time $t_i$ 
is located.
The usual lesser and greater projections are defined respectively as
$X^< \equiv X^{+-}$ and $X^> \equiv X^{-+}$, and the usual
time-ordered (anti-time-ordered) as $X^t=X^{++}$ ($X^{\tilde t}=X^{--}$).

By definition, complex conjugation of the different \GFs follows the rules:
\begin{equation}
\label{eq:app_cc}
\begin{split}
G^a(1,2)	& =   \left( G^r(2,1) \right)^*   \\
G^\gtrless(1,2)	& = - \left( G^\gtrless(2,1) \right)^* \nonumber
 \end{split}
\end{equation}
Similar expressions hold for the \SEs $\Sigma^x(1,2)$ .

\section{Rules for analytical continuation}
\label{app:analyticalcontinuation}

The rules for analytical continuation from $C_K$ to normal real-time make that 
the following products $P_{(i)}(\tau,\tau')$ on the time-loop contour,
\begin{equation}
\label{eq:app_acontinuation_CK}
\begin{split}
P_{(2)}	& =  \int_{C_K} A B   \\
P_{(3)}	& =  \int_{C_K} A B C  \\
P_{(n)}	& =  \int_{C_K} A_1 A_2 ... A_n, 
\end{split}
\end{equation}
have the following components $P_{(i)}^x(t,t')$ on the real-time axis $(x=r,a,>,<)$
\begin{equation}
\label{eq:app_acontinuation_t}
\begin{split}
P_{(2)}^\gtrless	& =  \int_t A^r B^\gtrless + A^\gtrless B^a   \\ 
P_{(3)}^\gtrless	& =  \int_t A^\gtrless B^a C^a + A^r B^\gtrless C^a + A^r B^r C^\gtrless  \\
P_{(n)}^r	& =  \int_t A_1^r A_2^r ... A_n^r  \hspace{7mm} 
P_{(n)}^a	  =  \int_t A_1^a A_2^a ... A_n^a.  
\end{split}
\end{equation}

\section{Derivation of the current $I_L$}
\label{app:derivationIL}

In this appendix, we provide the details of the derivation of the main results of this paper,
mainly Eq.~(\ref{eq:ILfinal}) and Eq.~(\ref{eq:Upsilons}).
In the following, we choose to use the symbol $\summline$ for summations in order to have a better 
graphical distinction between the sum-signs and the self-energies $\Sigma$.

From Eq.~(\ref{eq:ILafterEOM}) and Eq.~(\ref{eq:Trace_in_ILafterEOM_omega}), we need to calculate
the following traces:
\begin{equation}
\label{eq:Trace_LCR_SigmaG}
\begin{split}
{\rm Tr}_\lambda & \left[ (\Sigma G)^<  \right]
= \\
& \summ_{\lambda,n} \left( (V_{\lambda n}+ \Sigma^{\MB}_{\lambda n}) G_{n \lambda}\right)^<
+
\summ_{\lambda,\lambda'} (\Sigma^{\MB}_{\lambda \lambda'} G_{\lambda' \lambda})^< ,
\end{split}
\end{equation}
and similarly 
\begin{equation}
\label{eq:Trace_LCR_GSigma}
\begin{split}
{\rm Tr}_\lambda & \left[ (G \Sigma)^< \right]
= \\
& \summ_{\lambda,n} \left( G_{\lambda n} (\Sigma^{\MB}_{n \lambda} + V_{n \lambda}) \right)^<
+
\summ_{\lambda,\lambda'} (G_{\lambda \lambda'} \Sigma^{\MB}_{\lambda' \lambda})^<
\end{split}
\end{equation}
since
$\Sigma^{\MB}_{\lambda \rho}=0$ and $\Sigma^{\MB}_{\rho \lambda}=0$.

We first consider the sums $\summline_{\lambda,\lambda'}$
\begin{equation}
\label{eq:sums_lambdas}
\begin{split}
& \summ_{\lambda,\lambda'} (\Sigma^{\MB}_{\lambda \lambda'} G_{\lambda' \lambda})^<
-
\summ_{\lambda,\lambda'} (G_{\lambda \lambda'} \Sigma^{\MB}_{\lambda' \lambda})^<  = \\
& \summ_{\lambda,\lambda'} \Sigma^{\MB, <}_{\lambda \lambda'} G_{\lambda' \lambda}^a
+ \Sigma^{\MB,r}_{\lambda \lambda'} G_{\lambda' \lambda}^<
- G^<_{\lambda \lambda'} \Sigma^{\MB,a}_{\lambda' \lambda}
- G^r_{\lambda \lambda'} \Sigma^{\MB,<}_{\lambda' \lambda} \\
= & \summ_{\lambda,\lambda'} 
\Sigma^{\MB, <}_{\lambda \lambda'} ( G_{\lambda' \lambda}^a - G_{\lambda' \lambda}^r )
+ ( \Sigma^{\MB,r}_{\lambda \lambda'} - \Sigma^{\MB,a}_{\lambda' \lambda}) G_{\lambda' \lambda}^< \\
= & \summ_{\lambda,\lambda'} 
\Sigma^{\MB, <}_{\lambda \lambda'} (G^<-G^>)_{\lambda' \lambda}
+ ( \Sigma^{\MB,>} - \Sigma^{\MB,<} )_{\lambda \lambda'} G_{\lambda' \lambda}^< \\
= & \summ_{\lambda,\lambda'} 
\Sigma^{\MB, >}_{\lambda \lambda'} G^<_{\lambda' \lambda}
- \Sigma^{\MB,<}_{\lambda \lambda'} G_{\lambda' \lambda}^> \\
= & {\rm Tr}_{\lambda} \left[ \Sigma^{\MB, >}_L G^<_L - \Sigma^{\MB,<}_L G^>_L \right] .
\end{split}
\end{equation}

In the first line of Eq.~(\ref{eq:sums_lambdas}), 
we used the rules of analytical continuation. In the second, we have used
the equivalent of cyclic permutation in the calculation of a trace, i.e. swapping the index
$\lambda$ and $\lambda'$ in the last two terms. This is possible here since the sums and all
matrix elements are defined in the single subspace of the L electrode.
The final result looks like the collision terms usually obtained in the derivation of a generalised
Boltzmann equation from quantum kinetic theory. 
They correspond to the particle production (scattering-in) and absorption or hole production
(scattering-out) related to inelastic processes 
(i.e. non-diagonal elements of the \SE on the time-loop contour $\Sigma^<$) occuring in the left
electrode.

Now we consider the sums $\summline_{\lambda,n}$ in Eq.(\ref{eq:Trace_LCR_SigmaG}) and 
Eq.(\ref{eq:Trace_LCR_GSigma}).
We find that
\begin{equation}
\label{eq:app_IL_Tr_n_1}
\begin{split}
& \summ_{\lambda n} \left[ ... \right] = \\
& \summ_{\lambda n} \left[ 
  \Sigma^r_{\lambda n} G^<_{n \lambda} - G^<_{\lambda n} \Sigma^a_{n \lambda} 
+ \Sigma^{\MB <}_{\lambda n} G^a_{n \lambda} - G^r_{\lambda n} \Sigma^{\MB <}_{n \lambda} 
\right] .
\end{split}
\end{equation}

We now need to calculate the following different \GFs matrix elements
$G^<_{n\lambda}, G^<_{\lambda n}, G^a_{n\lambda}$ and $G^r_{\lambda n}$.
For this we use the Dyson-like equation defined for the non-diagonal elements:
$G^x_{n\lambda}=\langle n\vert (G\Sigma g)^x\vert\lambda\rangle$, and
$G^x_{\lambda n}=\langle \lambda\vert (G\Sigma g)^x\vert n \rangle$, and

We concentrate on one matrix element $\langle n\vert (G\Sigma g)^<\vert\lambda\rangle$
to show the mechanism of the derivation:

\begin{equation}
\label{eq:Gless_nlambda}
\begin{split}
 G^<_{n\lambda} & = \langle n\vert (G\Sigma g)^<\vert\lambda\rangle \\
& =\langle n\vert G^r \Sigma^< g^r + G^< \Sigma^a g^a + G^r \Sigma^r g^< \vert\lambda\rangle \\
& = \bullet\hspace{-6mm}\sum_{\lambda_1,\lambda_2,m} 
 G^r_{n\lambda_1}\ \Sigma^<_{\lambda_1 \lambda_2}\ g^a_{\lambda_2 \lambda}
+ G^r_{nm}\ \Sigma^<_{m \lambda_2}\ g^a_{\lambda_2 \lambda} \\
& \qquad  +  G^<_{n\lambda_1}\ \Sigma^a_{\lambda_1 \lambda_2}\ g^a_{\lambda_2 \lambda}
+ G^<_{nm}\ \Sigma^a_{m \lambda_2}\ g^a_{\lambda_2 \lambda} \\
& \qquad  +  G^r_{n\lambda_1}\ \Sigma^r_{\lambda_1 \lambda_2}\ g^<_{\lambda_2 \lambda}
+ G^r_{nm}\ \Sigma^r_{m \lambda_2}\ g^<_{\lambda_2 \lambda} ,
\end{split}
\end{equation}
with $\Sigma^{a/r}_{m \lambda}=V_{m \lambda}+\Sigma^{\MB,a/r}_{m \lambda}$ 
and $\Sigma^<_{m \lambda}=\Sigma^{\MB,<}_{m \lambda}$,
and, as explained above, we have used the condition $\Sigma^x_{\rho \lambda}=0$.
The same principle holds for the derivation of the other \GFs matrix elements.

The interesting point is that the terms in $\Sigma^{a/r}_{\lambda_1 \lambda_2}$ can
be factorised out and included within the renormalisation of the 
left lead \GFs $g^{a/r,<}_{\lambda_1 \lambda_2}$ as follows
\begin{equation}
\label{eq:def_tildega}
g^a_{\lambda \lambda'} \left( 1- \Sigma^{\MB,a} g^a \right)^{-1}_{\lambda' \lambda_1}
= \tilde{g}^a_{\lambda \lambda_1} .
\end{equation}
 
\vspace{3cm}

Therefore the matrix  $G^<_{n\lambda}=\langle n\vert (G\Sigma g)^<\vert\lambda\rangle$ can be recast as
$G^<_{n\lambda}=\langle n\vert (G_C\ \Sigma_{CL}\ \tilde{g}_L)^<\vert\lambda\rangle$, or similarly with
an explicit summation:
\begin{equation}
\label{eq:Gless_nlambda_bis}
\begin{split}
G^<_{n\lambda}= \summ_{m,\lambda'}
& G^r_{nm}\ \Sigma^r_{m \lambda'}\ \tilde{g}^<_{\lambda' \lambda} \\
+\ & G^r_{nm}\ \Sigma^{\MB,<}_{m \lambda'}\ \tilde{g}^a_{\lambda' \lambda}
+ G^<_{nm}\ \Sigma^a_{m \lambda_2}\ \tilde{g}^a_{\lambda' \lambda} .
\end{split}
\end{equation}

We also find that
\begin{equation}
\label{eq:app_GFmatrixelements}
\begin{split}
G^<_{\lambda n} & = \langle \lambda\vert (\tilde{g}_L\ \Sigma_{LC}\ G_C)^<\vert n \rangle , \\
G^a_{n \lambda} & = \langle n\vert (G_C\ \Sigma_{CL}\ \tilde{g}_L)^a\vert\lambda\rangle , \\
G^r_{\lambda n} & =\langle \lambda\vert (\tilde{g}_L\ \Sigma_{LC}\ G_C)^r\vert n \rangle .
\end{split}
\end{equation}

Using the rules of analytical continuation for products of three quantities, we find that 
Eq.(\ref{eq:app_IL_Tr_n_1}) becomes

\begin{widetext}

\begin{equation}
\label{eq:app_IL_Tr_n_2}
\begin{split}
& \summ_{\lambda n}
\langle\lambda\vert \Sigma^r_{LC} \vert n \rangle \ \langle n\vert (G_C\ \Sigma_{CL}\ \tilde{g}_L)^< \vert\lambda\rangle
- 
\langle \lambda\vert (\tilde{g}_L\ \Sigma_{LC}\ G_C)^<\vert n \rangle \ \langle n\vert\Sigma^a_{CL}\vert\lambda
\rangle
+
\langle\lambda\vert \Sigma^{\MB <}_{LC} \vert n\rangle \ \langle n\vert (G_C\ \Sigma_{CL}\ \tilde{g}_L)^a\vert\lambda\rangle \\
&
\qquad
- 
\langle \lambda\vert (\tilde{g}_L\ \Sigma_{LC}\ G_C)^r\vert n \rangle \ \langle n\vert\Sigma^{\MB <}_{CL} \vert\lambda\rangle  \\ \\
& = 
\summ_n 
\langle n \vert 
\left(
G^<_C\ \Sigma^a_{CL}\ \tilde{g}^a_L + G^r_C\ \Sigma^<_{CL}\ \tilde{g}^a_L  + G^r_C\ \Sigma^r_{CL}\ \tilde{g}^<_L 
\right)\Sigma^r_{LC}  \vert n \rangle 
+ \langle n \vert 
G^a_C\ \Sigma^a_{CL}\ \tilde{g}^a_L\ \Sigma^<_{LC}
- \Sigma^<_{LC}\ \tilde{g}^r_L\ \Sigma^r_{LC}\ G^r_C
\vert n \rangle \\
& 
\qquad 
+ \langle n \vert 
\Sigma^a_{CL}
\left(
\tilde{g}^<_L\ \Sigma^a_{LC}\ G^a_C + \tilde{g}^r_L\ \Sigma^<_{LC}\ G^a_C  + \tilde{g}^r_L\ \Sigma^r_{LC}\ G^<_C 
\right) \vert n \rangle \\ \\
& = 
\summ_n 
\langle n \vert 
  G^r_C \left( 
  \Sigma^<_{CL}\ \tilde{g}^a_L\ \Sigma^r_{LC} 
+ \Sigma^r_{CL}\ \tilde{g}^<_L\ \Sigma^r_{LC} 
- \Sigma^<_{CL}\ \tilde{g}^r_L\ \Sigma^r_{LC}  
\right)
+ G^a_C \left( 
  \Sigma^a_{CL}\ \tilde{g}^a_L\ \Sigma^<_{LC} 
- \Sigma^a_{CL}\ \tilde{g}^<_L\ \Sigma^a_{LC} 
- \Sigma^a_{CL}\ \tilde{g}^r_L\ \Sigma^<_{LC}  
\right) \\
& 
\qquad
+ G^<_C \left( \Sigma^a_{CL}\ \tilde{g}^a_L\ \Sigma^r_{LC} - \Sigma^a_{LC}\ \tilde{g}^r_L\ \Sigma^r_{LC} \right)
\vert n \rangle \\ \\
& =
\tr_n\left[ G^r_C \tilde{\Upsilon}^l_{LC} + G^a_C
  (\tilde{\Upsilon}^l_{LC})^\dagger  + G^<_C(\tilde\Upsilon_{LC} -
  \tilde{\Upsilon}_{LC}^\dagger) \right] .
\end{split}
\end{equation}
\end{widetext}
In the second equality of Eq.~(\ref{eq:app_IL_Tr_n_2}), 
the matrix elements have been swapped to get a trace only over the central region subspace $\{n\}$.
The final two equalities are just exactly the ${\rm Tr}_n$ entering the definition of the
current $I_L$ given by Eq.(\ref{eq:ILfinal}) with the definitions of 
$\tilde\Upsilon^l_{LC}$ and $\tilde\Upsilon_{LC}$
(and their adjoints) given by
\begin{equation}
\label{eq:Upsilons_app}
\begin{split}
\tilde\Upsilon^l_{LC} & = \Sigma^<_{CL} \left( \tilde{g}^a_L - \tilde{g}^r_L \right) \Sigma^r_{LC}
+ \Sigma^r_{CL}\ \tilde{g}^<_L\ \Sigma^r_{LC} \\
& = (\Sigma \tilde{g})^<_{CL}\  \Sigma^r_{LC}
- \Sigma^<_{CL}\ (\tilde{g} \Sigma)^r_{LC} , \\ \\
\tilde\Upsilon_{LC} & = \Sigma^a_{CL}\ \tilde{g}^a_L\ \Sigma^r_{LC} ,
\end{split}
\end{equation}
and 
\begin{equation}
\label{eq:Upsilons_dag_app}
\begin{split}
(\tilde\Upsilon^l_{LC})^\dag & = \Sigma^a_{CL} \left( \tilde{g}^a_L - \tilde{g}^r_L \right) \Sigma^<_{LC}
- \Sigma^a_{CL}\ \tilde{g}^<_L\ \Sigma^a_{LC} \\
& = (\Sigma \tilde{g})^a_{CL}\  \Sigma^<_{LC}
- \Sigma^a_{CL}\ (\tilde{g} \Sigma)^<_{LC} , \\ \\
\tilde\Upsilon^\dag_{LC} & = \Sigma^a_{CL}\ \tilde{g}^r_L\ \Sigma^r_{LC} .
\end{split}
\end{equation}
(QED) Eqs.~(\ref{eq:sums_lambdas},\ref{eq:app_IL_Tr_n_2},\ref{eq:Upsilons_app},\ref{eq:Upsilons_dag_app}) are just
the main results of this paper. 

Now we can follow the same mechanism of derivation to obtain an expression similar to Eq.~(\ref{eq:ILfinal}) 
for the current $I_R$ flowing at the right {\em CR} interface. In concrete the expression for $I_R$ is 
given by Eq.~(\ref{eq:ILfinal}) by swapping the index $L$~$\leftrightarrow$~$R$ and with a minus sign 
because of the current conservation condition $I_L+I_R=0$.

Finally one should note that because of the following three conditions:
{\em (i)} the very existence of the interaction crossing at the contact,
{\em (ii)} the fact that in the most general cases
$\Sigma^a_{C\alpha / \alpha C} \ne \Sigma^r_{C\alpha / \alpha C}$, in opposition with the 
non-interaction case where 
$V^a_{C\alpha/\alpha C} = V^r_{C\alpha / \alpha C} = V_{C\alpha / \alpha C}$,
and {\em (iii)} the rules of analytical continuation for triple products $P_{(3)}$,
the usual cyclic permutation performed in the calculation of the trace 
${\rm Tr}_\lambda[(\Sigma G)^<-(G \Sigma)^<]$ 
cannot be used to transform the initial trace over $\{\lambda\}$ onto a trace over $\{n\}$. 
Therefore
the current $I_L$ at the $LC$ contact is not given by a straightforward generalisation 
of the Meir and Wingreen formula of the type
\begin{equation}
\begin{split}
I_L \ne \frac{e}{\hbar} \int \frac{d\omega}{2\pi} 
& {\rm Tr}_n [Y^<_L G^>_C - Y^>_L G^<_C] \\
&+{\rm Tr}_\lambda [\Sigma^{\MB >}_L G^<_L - \Sigma^{\MB <}_L G^>_L].
\end{split}
\end{equation}
where $Y^x_L$ is the generalised (interacting) embedding potential of the $L$ electrode.

This is another very important result of our work which has strong implication in the expression of the current itself, and also in the conditions of current conservation.

\section{Current conservation condition}
\label{app:currentconserv}

First we consider the general definition of the lesser and greater \GFs:

\begin{equation}
\label{eq:app_defGlessgrt_1}
G^\gtrless = 
( 1 + G^r \Sigma^r ) g^\gtrless ( 1 + \Sigma^a\ G^a ) + G^r\ \Sigma^<\ G^a .
\end{equation}
The first term represents the initial conditions $g^\gtrless$.

For the central region, we have chosen the initial condition such as 
$\langle n\vert g^< \vert m\rangle=0$ (see Eq.(\ref{eq:GlessC})). We could have
chosen another initial condition. Such choices have no effects on the steady
state regime when a steady current flow through the central region, however
the initial conditions play an important role in the transient behaviour of
the current \cite{Tran:2008,Myohanen:2008,Perfetto:2010,Velicky:2010}.

For the definition of the lesser left and right Green's functions, it is however not possible
to neglect the initial conditions (before full interactions and coupling to the
region central are taken into account).
This is because it would not be physical to ignore
the presence of the left and right Fermi seas, obtained as the thermodynamical
limit of the two semi-infinite leads which act as electron emitter and collector
in our model device.

One can however recast Eq.(\ref{eq:app_defGlessgrt_1}) as follows
\begin{equation}
\label{eq:app_defGlessgrt_2}
G^< = G^r\ ( (g^r)^{-1} g^< (g^a)^{-1} + \Sigma^{<} )\ G^a
= G^r\ \bar{\Sigma}^{<}\ G^a
\end{equation}
with $\bar{\Sigma}^{<} = \Sigma^{<} + \gamma^<$ and $\gamma^< = (g^r)^{-1} g^< (g^a)^{-1}$.
And similarly for $G^>$.
Hence $\gamma^< - \gamma^> = (g^a)^{-1} - (g^r)^{-1}$ and
$\bar{\Sigma}^< - \bar{\Sigma}^> = (G^a)^{-1} - (G^r)^{-1}$.

From these properties, it can be easily shown that 
\begin{equation}
\label{eq:app_collisionTerm_Tr_all}
{\rm Tr}_{\rm all} \left[ \bar\Sigma^< G^> - \bar\Sigma^> G^< \right] = 0
\end{equation}
for each $\omega$. The trace runs over all indexes in the system (${\rm all} \equiv \{ \lambda, n, \rho \}$)
and the interaction $\Sigma$ are spread over the whole $L,C,R$ regions. This is the starting point
to find the conditions for current conservation.

Because the trace runs over all the three subspaces, we can apply the usual cyclic permutation and recast
Eq.(\ref{eq:app_collisionTerm_Tr_all}) as follows
\begin{equation}
\label{eq:app_collisionTerm_Tr_all_bis}
-{\rm Tr}_{\rm all} \left[ (\Sigma G)^< - (G \Sigma)^< \right] +
{\rm Tr}_{\rm all} \left[ \gamma^< G^> - \gamma^> G^< \right] = 0
\end{equation}
or equivalently
\begin{equation}
\label{eq:app_collisionTerm_Tr_all_ter}
\int {\rm d}\omega\ {\rm Tr}_{\rm all} \left[ (\Sigma G)^< - (G \Sigma)^< \right] +
{\rm Tr}_{\rm all} \left[ \gamma^> G^< - \gamma^< G^> \right] = 0
\end{equation}

Expanding the trace in the first term over each subspace 
${\rm Tr}_{\rm all}[...] = {\rm Tr}_{\lambda}[..]+{\rm Tr}_{n}[...]+{\rm Tr}_{\rho}[...]$,
one can identify the definition of the currents $I_L$ and $I_R$ from ${\rm Tr}_{\lambda}[...]$
and ${\rm Tr}_{\rho}[...]$ respectively.

Hence the condition of current conservation $I_L+I_R=0$ leads to
\begin{equation}
\label{eq:app_curconserv_cond_1}
\int {\rm d}\omega\ {\rm Tr}_{\rm n} \left[ (\Sigma G)^< - (G \Sigma)^< \right] +
{\rm Tr}_{\rm all} \left[ \gamma^> G^< - \gamma^< G^> \right] = 0 .
\end{equation}

After further manipulation, lengthy but trivial in the light of Appendix \ref{app:derivationIL},
we find that the current conservation leads to the following condition:
\begin{equation}
\label{eq:currentconserv}
  \int {\rm d}\omega \
   {\rm Tr}_{\alpha=L,C,R} [ \Sigma^{\MB >}_\alpha G^<_\alpha  - \Sigma^{\MB <}_\alpha G^>_\alpha ] + 
C_{LC}+C_{CR}=0
\end{equation}
with
\begin{equation}
\begin{split}
C_{LC}(\omega)={\rm Tr}_{n}  [ (& \tilde Y_L^> + \tilde\Upsilon_{LC} -
    \tilde\Upsilon^\dag_{LC}- \tilde\Upsilon^l_{LC}) G^<_C \\ 
   - (& \tilde Y_L^< - \tilde\Upsilon^l_{LC}) G^>_C +
    (\tilde\Upsilon^l_{LC}+(\tilde\Upsilon^l_{LC})^\dag) G^a ]
\end{split}
\end{equation}
and $C_{CR}=C_{LC}[ \{L \leftrightarrow R \}]$.

The first trace in Eq.~(\ref{eq:currentconserv}) corresponds to the sum over the three 
regions $L,C,R$ of the integrated collision term.  The two other traces $C_{LC}$ and $C_{CR}$
arise from the interactions crossing at the $LC$ and $CR$ interfaces.  But, globally, 
Eq.~(\ref{eq:currentconserv}) still implies that the total integrated collision terms must vanish.


\begin{thebibliography}{35}
\bibitem{footnote_SigmaLR}
The screening processes are such that either an electron in the region $L$ cannot directly interact
with another 
electron in the region $R$ or such interaction is included into 
the $\Sigma^{\rm MB}_{\lambda m}$ and $\Sigma^{\rm MB}_{\rho n}$ components.
\end{thebibliography}

\end{document}